%% file: main.tex
\documentclass[conference]{IEEEtran}
\IEEEoverridecommandlockouts
\usepackage{cite}
\usepackage{amsmath,amssymb,amsfonts}
\usepackage{algorithmic}
\usepackage{graphicx}
\usepackage{textcomp}
\usepackage{xcolor}

\usepackage{caption}
\usepackage{subcaption}

\def\BibTeX{{\rm B\kern-.05em{\sc i\kern-.025em b}\kern-.08em
    T\kern-.1667em\lower.7ex\hbox{E}\kern-.125emX}}
    
\makeatletter
\newcommand{\linebreakand}{%
  \end{@IEEEauthorhalign}
  \hfill\mbox{}\par
  \mbox{}\hfill\begin{@IEEEauthorhalign}
}
\makeatother    
    
\begin{document}

\title{VSDM: A Virtual Service Device Management Scheme for UPnP-Based IoT Networks}

\author{\IEEEauthorblockN{Golam Kayas}
\IEEEauthorblockA{\textit{Dept. of Computer \& Info. Science} \\
\textit{Temple University, USA}\\
golamkayas@temple.edu}
\and
\IEEEauthorblockN{Mahmud Hossain}
\IEEEauthorblockA{\textit{Dept. of Computer Science} \\
\textit{University of Alabama at Birmingham, USA}\\
mahmud@uab.edu}
\and
\IEEEauthorblockN{Jamie Payton}
\IEEEauthorblockA{\textit{Dept. of Computer \& Info. Science} \\
\textit{Temple University, USA}\\
payton@temple.edu}
\linebreakand 
\IEEEauthorblockN{ S. M. Riazul Islam}
\IEEEauthorblockA{\textit{Dept. of Computer Engineering} \\
\textit{Sejong University, South Korea}\\
riaz@sejong.ac.kr}

}

\newcommand{\model}{ VSDM }
\IEEEoverridecommandlockouts
\IEEEpubid{\makebox[\columnwidth]{978-1-7281-9656-5/20/\$31.00~\copyright2020 IEEE \hfill} \hspace{\columnsep}\makebox[\columnwidth]{ }}
\maketitle
\IEEEpubidadjcol

\begin{abstract}
\input{abstract}
\end{abstract}

\input{introduction}
\input{background}

\input{proposed-model}
\input{experiment}

\input{discussion-realted-work}
\input{conclusion-future-work}
\input{acknowldgement}

\bibliographystyle{IEEEtran} 
\bibliography{bibliography.bib}
\end{document}

%% file: abstract.tex
The ubiquitous nature of IoT devices has brought new and exciting applications in computing and communication paradigms. Due to its ability to enable auto-configurable communication between IoT devices, pervasive applications, and remote clients, the use of the Universal Plug and Play (UPnP) protocol is widespread. However, the advertisement and discovery mechanism of UPnP incurs significant overhead on resource-constrained IoT devices. In this paper, we propose a delegation-based approach that extends the UPnP protocol by offloading the service advertisement and discovery-related overhead from resource-limited IoT devices to the resource-rich neighbours of a UPnP-enabled IoT network. Our experimental evaluations demonstrate that the proposed scheme shows significant improvement over the basic UPnP, reducing energy consumption and network overhead. 

%% file: introduction.tex
\section{Introduction}
The past decade has seen substantial growth in the manufacturing of Internet of Things (IoT) devices. As the availability of a range of different types of sensor- and actuator-enabled IoT devices at the network's edge has increased, a number of IoT deployments have emerged, including smart home systems \cite{chowdhury2019iot} that integrate home security~\cite{9139467}, management, and convenience for home owners; health and wellness services \cite{qadri2020future, russell2019probabilistic, hossain2018internet} that allow for remote, continuous, multi-modal monitoring of physiological and behavioral characteristics~\cite{samie2020hierarchical}; and intelligent agriculture systems~\cite{glaroudis2020survey} that can optimize crop growth and harvesting. Common to these and future IoT application deployments is the need for an open system  that supports fluid, dynamic, opportunistic integration of potentially mobile clients and services supported by IoT devices at scale, with minimal configuration.
%connected more IoT devices in the pervasive networks everyday. IoT devices offer advance connectivity and useful applications to a wide varity of smart systems. In the IoT ecosystem, usually a lots of different kinds of sensors and devices are co-located and communicate with each other to achieve a certain goal. But the heterogeneity of devices in a IoT network is subjected to interoperability challenges. Moreover, IoT devices are often resource constrained in nature, which makes the adaptation of the traditional networking solutions harder and infeasible. 

Given these needs, service-oriented architectures are well-suited to support IoT-enabled systems, providing the ability for resource-constrained IoT devices to advertise software services that can be discovered and used by applications. The Universal Plug and Play (UPnP) \cite{upnp} protocol, which implements a service-oriented model and can be used to connect clients and services across a network, has been widely adopted for use to support IoT systems. UPnP supports dynamic service advertisement and service discovery, and supports zero configuration networking. It also offers language independence and interoperability, supporting the incorporation of a wide variety of devices from different manufacturers with varying capabilities and configurations. 

However, supporting UPnP on the types of small, resource-constrained devices that are common in IoT networks is not without challenges. In particular, despite recent advances in device and network protocol design, the substantial power consumption required to send and receive messages via a wireless medium makes communication overhead a significant concern for small, battery-powered sensor devices. Specifically, in the UPnP context, an IoT device acting as a service device (SD) typically needs to periodically multicast advertisements to announce the availability of the services it provides. The advertisement message also contains the information needed to access the services provided by the SD. As the IoT devices are resource-constrained and usually battery-powered, this periodic multicast advertisement message incurs considerable energy overhead. 

In this work, we build on the observation that resource-constrained IoT devices at the edge of the network are often connected to a more resource-rich  neighbours, such as a gateway device. In general, a gateway device act as a bridge between heterogeneous IoT devices within the local area network and remote devices, performing protocol translation, packet fragmentation, and communication bridging between different communication technologies (e.g., BLE \cite{ble}, Zigbee ~\cite{zigbee}, 6LoWPAN~\cite{6lowpan}). With respect to supporting UPnP in IoT networks, the gateway can also serve to assist resource-constrained IoT service devices by supporting delegation of service advertisements. Essentially, the gateway device can act as a {\em Virtual Service Device (VSD)}, which emulates the advertisement behavior of an SD; it handles sending the service advertisement and serves the service discovery requests from interested clients. This approach can reduce the network overhead, and therefore the energy consumption, incurred on IoT service devices in the UPnP-enabled network. 

In this work, we extend the UPnP protocol to incorporate service advertisement delegation from IoT service devices to a resource-rich Virtual Service Device (VSD) and propose a Virtual Service Device Management (VSDM) scheme.  The contributions of this work are summarized as follows:
\begin{itemize}
    \item We identify the energy consumption of UPnP service devices to advertise the targeted services of an IoT network.
    \item We propose a Virtual Service Device Management (VSDM) scheme, delegating the service advertisement overheads to a resource-rich network member for resource constrained IoT devices.
    \item We implement a prototype of VSDM and conduct an experimental evaluation. Our experiments focus on exploring the energy consumption and network overheads in basic UPnP and VSDM-enabled UPnP in an both IoT network.
\end{itemize}

%% file: background.tex
\vspace{-10pt}
\section{Background}
% \subsection{UPnP for IoT}

\subsection{UPnP based IoT Systems}
%\textcolor{blue}{JP: this subsection seems like it could be integrated into the next section (proposed scheme)? This section mostly seems to set up the protocol, and the part about the hardware/network architecture seems out of place and disconnected. Also, are we really referring to anything useful in Figure 2b that couldn't be combined with figure 3b?}

%Figure~\ref{fig:upnp-network} shows an example UPnP network with IoT devices. In a UPnP network, devices can be located in different types of networks such as BLE, Zigbee,  LoWPAN. The participants from different communication mediums can interact with each other to perform UPnP operations. For example, a smart phone that uses WiFi can act as a CP to discover a service provided by an IoT device in the 6LoWPAN network. The gateway device in responsible in bridging different communication technologies.  

\begin{figure}[h]
	\centering
	 \includegraphics[width=\columnwidth]{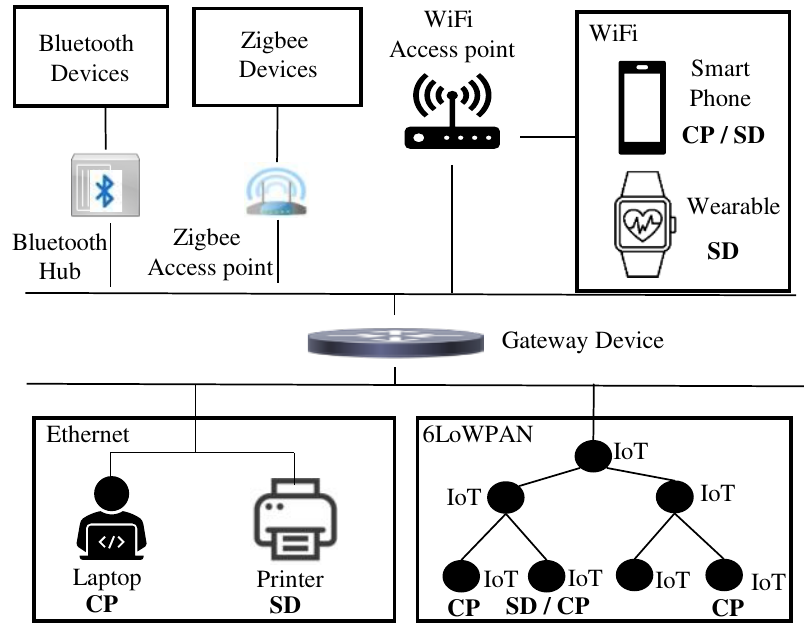}
  \caption{A UPnP Network with IoT devices.}
  \label{fig:upnp-network}
\end{figure}

Figure~\ref{fig:upnp-network} shows an example UPnP network with IoT devices. In a UPnP network, devices can be located in different types of networks such as BLE, Zigbee, and 6LoWPAN. The participants from different communication mediums can interact with each other to perform UPnP operations. For example, a smart phone that uses WiFi can act as a CP to discover a service provided by an IoT device in the 6LoWPAN network. The gateway device in responsible in bridging different communication technologies.  

\subsection{UPnP Components}
The components of a UPnP based system are classified into three categories: Service Device (SD), Control Point (CP), and Service.

\subsubsection{Service Device (SD)} In UPnP, an SD functions as a server that provides useful services to the clients upon request. In IoT-based scenarios, SDs are embedded with sensors that collect contextual information, actuators that can perform actions in response to sensed information, and radio transceivers for communication, with the support of a real-time operating system. For example, a service device may be a smart refrigerator, doorlock, or security camera in a smart home network.

\subsubsection{Control Point (CP)} A CP acts as a client that requests and consumes the services provided by the SDs. A CP can discover the available services, control the services, and request updates on the state change of the services. 

\subsubsection{Service} A service is a unit of functionality implemented by an SD located on the edge of the network. For example, a smart refrigerator device may offer a service to check the temperature of the vegetable drawer or to add an item to a shopping list.

\subsection{UPnP Phases and Operations}

\begin{figure}[t] 
    \centering
    \includegraphics[width=\columnwidth]{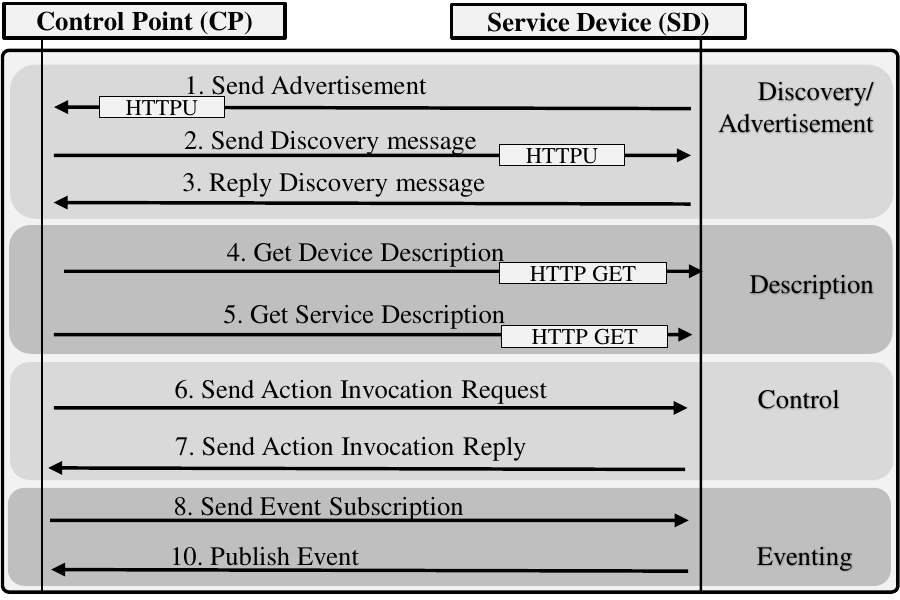}
  \caption{Interaction between a CP and an SD.}
  \label{fig:upnp-interaction}
\end{figure}

\begin{figure}[t] 
 \includegraphics[width=\columnwidth]{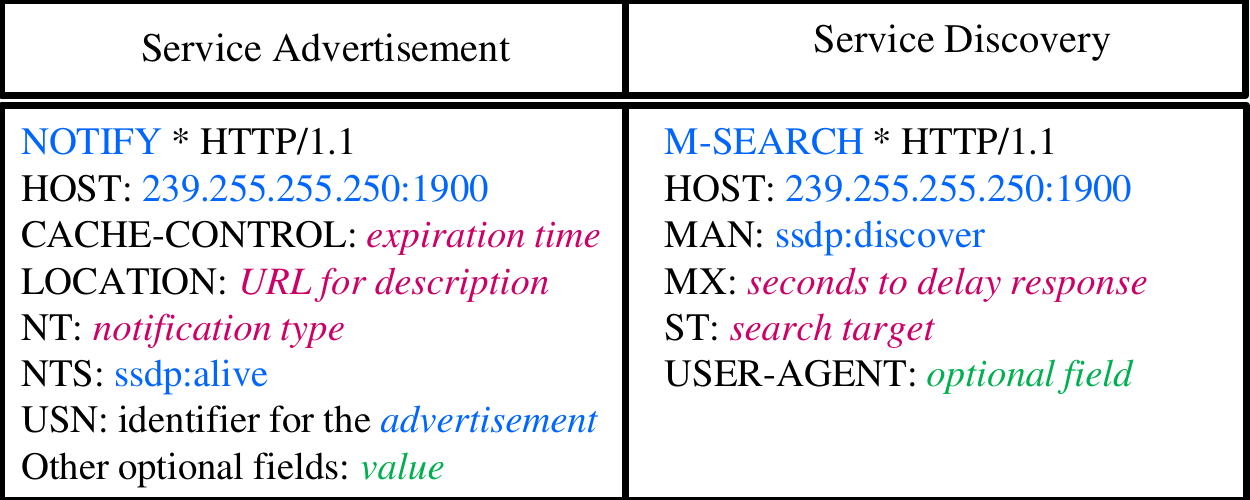}
  \begin{center}
  \caption{UPnP Advertisement and Discovery messages.}
  \label{fig:upnp-msg}
  \end{center}
\end{figure}

Figure~\ref{fig:upnp-interaction} shows the operations performed in UPnP communication by the SDs and the CPs in different phases. As shown in the figure, there are four essential phases of the UPnP protocol: Advertisement, Discovery, Description, Control, and Eventing.

\subsubsection{Advertisement} In the UPnP protocol~\cite{upnp-arch-2.0}, an SD periodically sends advertisement messages to the network by multi-casting to a standard address and port. Figure~\ref{fig:upnp-msg} shows an advertisement message sent by an SD. In the advertisement message (Figure~\ref{fig:upnp-msg}), the SD provides a URI (via the \texttt{LOCATION} field) that allows the CP to retrieve additional information about the device and its services. An SD will also unicast a similar message in response to a discovery message received from a CP. 

\subsubsection{Discovery} A CP searches a desired service in the network by broadcasting a discovery message. Figure~\ref{fig:upnp-msg} shows a sample discovery message sent by a CP. A CP defines the target service of the discovery message using the \texttt{ST} field (See Figure~\ref{fig:upnp-msg}). In reply of a discovery message, the CP receives a message similar to the advertisement revealing the description of the services.

\subsubsection{Description} After the advertisement and discovery, a CP only knows a URI location that provides the details of the services by a particular SD. In the description phase, the CP requests the URI location to retrieve the device description of the SD.

\subsubsection{Control} After retrieving the description of the services in the previous phase, a CP knows the name of the actions supported by the service, their parameters, and the way to invoke them. The CP sends control request to the service to perform the targeted actions.

\subsubsection{Eventing} Additionally, a CP also knows all the state variables of a service from its description and can subscribe to monitor the state variables. A subscription is when the targeted state variable of the specific services changes, every subscriber (in this case the CP) gets a notification from the SD.

For IoT devices, periodic multicasting of advertisement message can be prohibitively costly in terms of energy consumption due to communication overhead. Additionally, IoT service devices are vulnerable to service discovery flooding attacks, in which a malicious CP repeatedly sends service discovery messages; at a minimum, the IoT service device will incur significant communication overhead (and therefore, energy consumption) in responding to high volumes of service discovery requests and the IoT service device can be overwhelmed by the requests, effectively resulting in a a denial of service. We discuss these limitations of UPnP service advertisement in further detail in Section~\ref{sec:proposed} to motivate our proposed scheme.

%% file: proposed-model.tex
\section{Proposed Scheme: \model{}}\label{sec:proposed}

%\subsection{Problem Statement}
To support UPnP in IoT networks, an IoT SD  multicasts periodic advertisement messages, which consumes significant energy for a resource-constrained device. The IoT SD must also reply to discovery requests of the CPs. As such, IoT service devices are vulnerable to large volumes of service requests, whether originating from legitimate control points or malicious actors. In any case, so called ``discovery flooding'' incurs significant communication overhead on the IoT SD, which must respond to the high volumes of service discovery requests, and can ultimately exhaust the energy in a battery-powered IoT device. The IoT service device's buffer and processor can also be overwhelmed by the requests, effectively resulting in a different type of a denial of service, impacting the performance of the IoT system. Moreover, IoT devices are often deployed in low power and lossy networks (LLN) like 6LoWPAN. LLN networks are very constrained, and, in case of discovery flooding, these networks become congested, resulting in yet another type of performance impact. 

In response to aforementioned  limitations of deploying UPnP service advertisement in IoT networks, we propose a Virtual Service Device Management (VSDM) scheme to offload the advertisement and discovery reply to a resource-rich device in the IoT network. In VSDM scheme, the periodical advertisement and discovery reply is delegated to a {\em virtual service device (VSD)} which is a resource-rich member of the UPnP enabled IoT network. The VSDs take care of the advertisement and discovery reply, offloading the resource constrained IoT SDs. For later UPnP phases after discovery and advertisement, such as action request (control) and event subscription (eventing), the VSD redirects the requests to the SD. In this work, we focus on presenting a solution that extends the UPnP protocol to incorporate a single VSD in an UPnP-enabled IoT networks. However, we note that multiple VSDs can be deployed in a VSDM-enabled UPnP solution for IoT networks. Such a solution can offer additional benefits, including load balancing across VSDs and fault tolerance.

\begin{figure}[t] 
 \centering
     \includegraphics[width=\columnwidth]{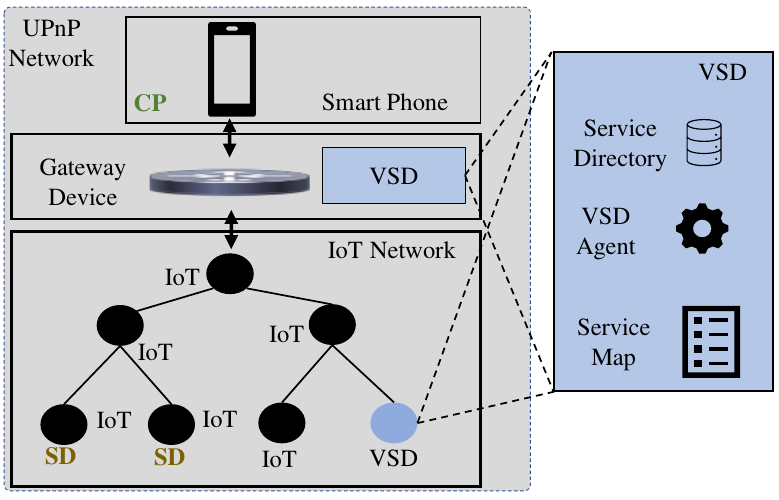}

 \caption{VSDM System Components.}
  \label{fig:vsdm-comp}
\end{figure}

\subsection{\model{} Components}
 
 Figure~\ref{fig:vsdm-comp} shows the system components of VSDM scheme. The \texttt{Virtual Service Device} (VSD) is implemented in the gateway device or in a resource-rich device in the same communication network. As most IoT networks are accompanied by relatively resource rich gateway devices, they are excellent candidates to implement a VSD. However, the gateway device is not the only candidate to become a VSD; any other participants that have enough resources can eventually implement the characteristics of the VSD. Every VSD implements a service named \texttt{VSD-Agent}, keeps a \texttt{Service Directory}, and maintains \texttt{Service Map}.

\subsubsection{VSD-Agent} The VSD-Agent is the service implemented by every VSD to support delegation. The description of the VSD-Agent service is maintained in a service description document. Figure~\ref{fig:vsd-des-doc} shows the JSON envelope of the service description document of the VSD-Agent service. This description document provides the URL to send action invocation requests, and exposes the name and arguments of the actions to be invoked. There are four actions implemented by the VSD-Agent service:
\begin{itemize}
    \item \textbf{Service-add:} \texttt{Service-add} is invoked by a SD when it tries to enroll itself to the VSD for delegation or it has a new service to provide. This action takes a list of service information as a parameter. Each entry of the service information list is a tuple like, \texttt{<service-name, service-type, description-location-url, control-url, event-url>}.  \texttt{Service-add} retrieves the service description from the original SDs and stores them in the Service Directory. It also inserts an entry to the Service-map mapping the delegation information.
    \item \textbf{Service-remove:} \texttt{Service-remove} is invoked by a SD when a previously advertised service is no longer available or the SD is leaving the UPnP network. This action also takes a list of service information as a parameter. \texttt{Service-remove} removes the delegated service description stored in the Service Directory and the Service-map entries for the associated services.
    \item \textbf{Service-update:} When a SD has an update for exiting service, it invokes the \texttt{Service-update} action. This action retrieves the new service description document from the SD and updates the Service-Directory and Service-map with the updated information.
    \item \textbf{Discovery-reply:} This action is invoked by the VSD that implements the VSD-Agent service upon receiving a discovery request from a CP. \texttt{Discovery-reply} takes the name of the targeted service and the address of the CP that issued the discovery request as a parameter. This action finds the targeted service name from the Service Map, constructs a delegated discovery reply, and sends the discovery reply to the CP.
    \item \textbf{Multicast-Advertisement :} This action is also invoked by the VSD periodically multicasting the advertisement for the available services. \texttt{Discovery-reply} iterates through the Service Map, constructs delegated advertisement messages for each entry, and  multicasts the advertisements in the network.       

\end{itemize}

\subsubsection{Service Directory} VSD retrieves the service descriptions of the SDs and stores them into the Service Directory to perform the delegation. While advertising the service provided by an SD, VSD uses the corresponding service description stored in the Service Directory as the delegated service description location.

\subsubsection{Service-Map} 
\begin{figure}[t] 
 \includegraphics[width=\columnwidth]{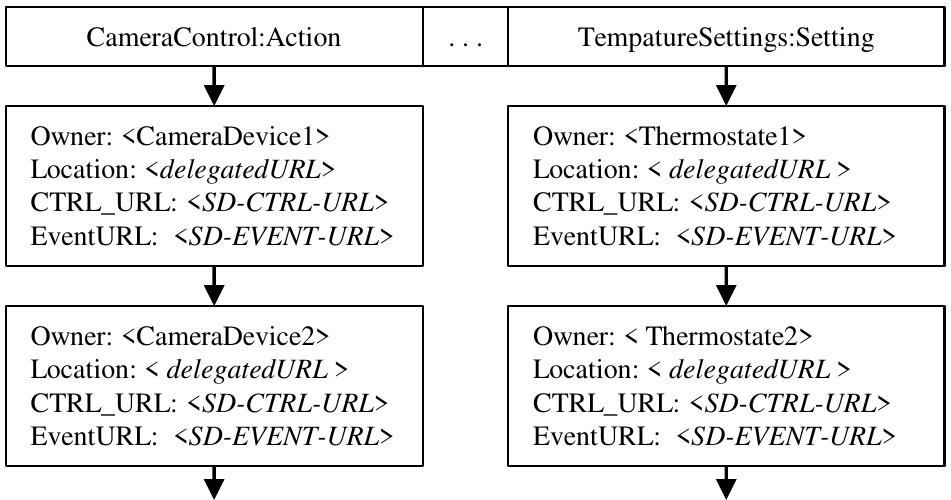}
  \begin{center}
  \caption{Structure of the Service Map.}
  \label{fig:svc-map}
  \end{center}
\end{figure}

Service-Map keeps the mapping of a service name and type to the description, control and event URLS and the Owner SD that provides the original service. 
The Service Map is implemented as a hashtable with separate chaining. 
Figure~\ref{fig:svc-map} shows an example of the \texttt{Service Map} data structure. 
There the description location (\texttt{Location}) is a delegated URL pointed to a URL stored in Service Directory. 
The control URL (\texttt{CTRL\_URL}) and the event URL are the absolute URL's provided by the SD.

\begin{figure}[t] 
 \centering
     \includegraphics[width=\columnwidth]{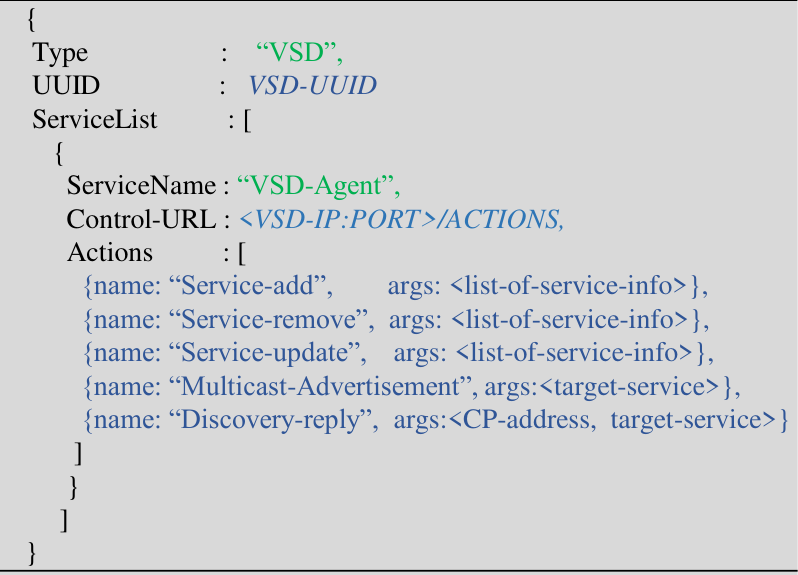}

 \caption{The JSON envelope of the VSD service description document.}
  \label{fig:vsd-des-doc}
\end{figure}

\begin{figure}[h] 
 \includegraphics[width=\columnwidth]{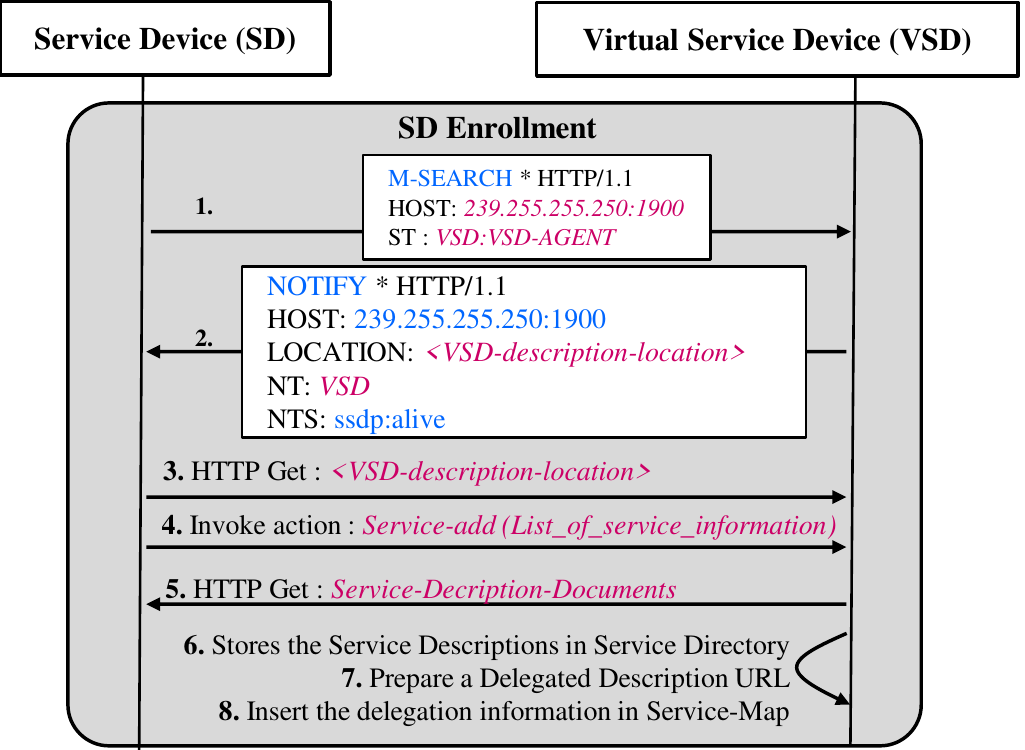}
  \begin{center}
  \caption{Enrollment of an SD in VSDM.}
  \label{fig:sd-enroll}
  \end{center}
\end{figure}
% \begin{figure}[t] 
%  \centering
%   \includegraphics[width=\columnwidth]{./figures/vsdm-interaction.pdf}
%  \caption{VSDM Interactions.}
%   \label{fig:vsdm-interaction}
% \end{figure}

\subsection{SD Enrollment}
Figure~\ref{fig:sd-enroll} shows the enrollment process for a SD.
\textbf{Step 1:} When a new SD enters the UPnP network, it multicasts a VSD discovery message in the network to find the VGD-Agent service. 
The VSD discovery message is based on the \texttt{ SSDP M-SEARCH} message like UPnP discovery, where the search target (\texttt{ST}) is  (\texttt{VSD:VSD-AGENT}). 
\textbf{Step 2:} The VSD device replies to the SD with the description location of the VSD-Agent service (\texttt{LOCATION} field).
\textbf{Step 3:} The SD sends a \texttt{HTTP GET} request to retrieve the description of the VSD-Agent service (see Figure~\ref{fig:vsd-des-doc}).
\textbf{Step 4:} The SD invokes the \texttt{Service-add} action of the VSD-Agent service with a list of the service information it wants to advertise in the network.
\textbf{Step 5: } The VSD-Agent sends \texttt{HTTP GET}
request to retrieve the service description documents of the services hosted by the SD.
\textbf{Step 6-8: } After receiving the service documents of the SD, VSD-Agent stores them it the Service Directory and creates a delegated description URL. Then VSD-Agent inserts an entry to the Service-Map with the service name, type, delegated description URL, control and the event URL of the SD  for each service.
\vspace{-5pt}
\subsection{Delegated Discovery Reply and Description}
\begin{figure}[t] 
 \includegraphics[width=\columnwidth]{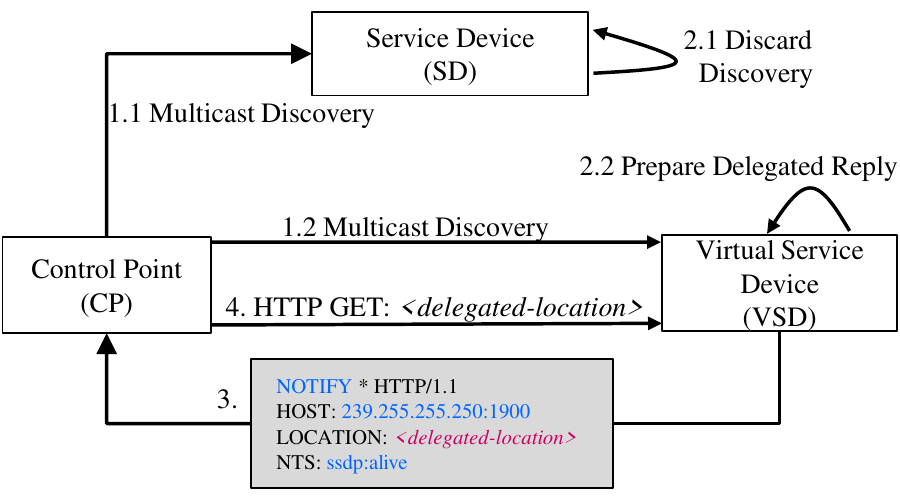}
  \begin{center}
  \caption{Delegated Discovery and Description of VSDM.}
  \label{fig:vsdm-dis}
  \end{center}
\end{figure}

Figure~\ref{fig:vsdm-dis} shows the delegated discovery and the description phases of VSDM. \textbf{Step 1:} The CP multicasts a discovery message in the network. \textbf{Step 2:} As the discovery message is sent using multicast, both the SD and the VSD receive it. As the SD delegates the discovery reply to the VSD, it ignores the discovery message. \textbf{Step 3:}  The VDS invokes the \texttt{Discovery-reply} action of the VSD-Agent and prepares a discovery reply using the delegated \texttt{LOCATION} URL. \textbf{Step 4:} The CP uses the delegated description location to get the service description documents from the VSD.

\subsection{VDSM Control and Eventing}
\begin{figure}[t] 
 \includegraphics[width=\columnwidth]{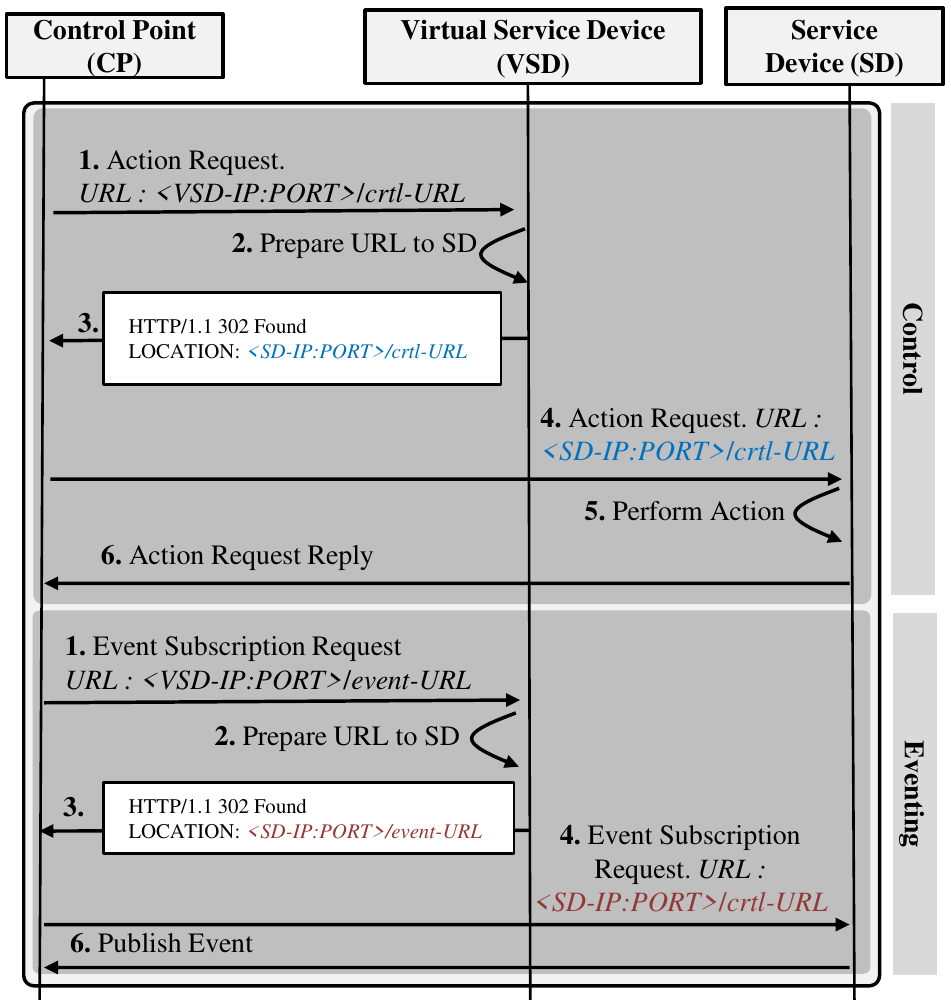}
  \begin{center}
  \caption{VSDM Control and Eventing Phase.}
  \label{fig:vsdm-crtl-event}
  \end{center}
\end{figure}

Figure~\ref{fig:vsdm-crtl-event} shows the control and eventing phases of the VDSM.
After getting the description of the services from the VSD, a CP performs action request invocation.  As the CP gets the description documents from the VSD, naturally it appends the address of the VSD with the relative control URL provided in the service description to invoke action invocation request (Step 1 Figure~\ref{fig:vsdm-crtl-event}). The VSD retrieves the absolute address of the control URL to the original SD from the Service-Map (see Figure~\ref{fig:svc-map}). 
Then the VSD redirects the control request to the original SD using \texttt{HTTP 302} redirection (Step 3 in Figure~\ref{fig:vsdm-crtl-event}). Thus the action invocation request of the CP is directly served by the original SD.

Similarly, the event subscription request is also redirected by the VSD to the original SD to publish the events to the CP.

%% file: experiment.tex
\section{Experiment and Evaluation}
\subsection{Experimental Setup}

\begin{figure}[t]
 \includegraphics[width=\columnwidth]{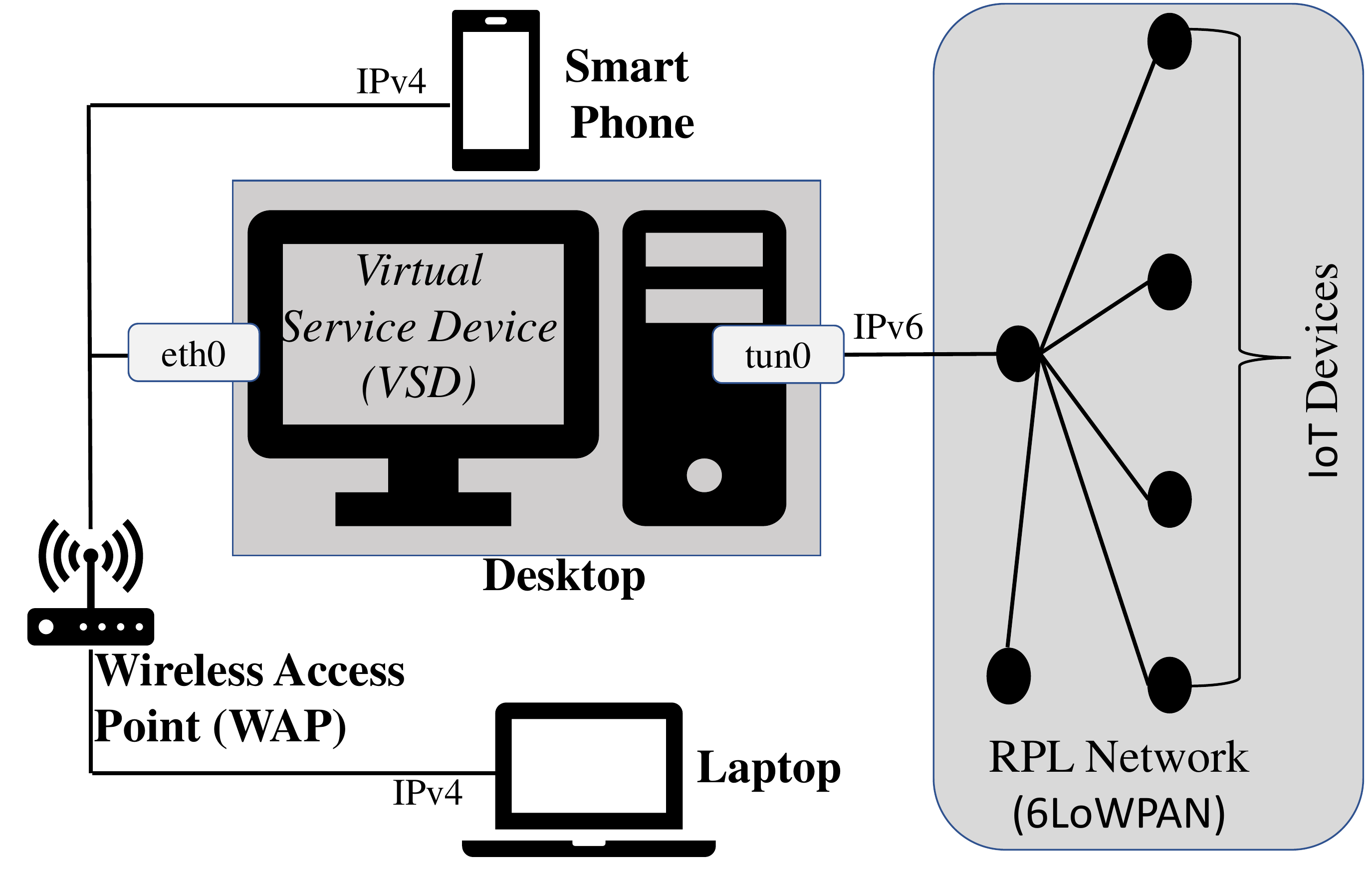}
  \begin{center}
  \caption{Experimental Network of\model.}
  \label{fig:exp-net}
  \end{center}
\end{figure}
The experimental UPnP network imitating a real-world UPnP enabled IoT setting is illustrated in Figure~\ref{fig:exp-net}. We used  Contiki operating system based simulator Cooja \cite{cooja}. We build an IoT device network that consists of some physical device and a number simulated devices. The simulated IoT  network, includes  a  number  of  Z1~\cite{z1} mote devices,  with communications supported by 6LoWPAN  and the RPL~\cite{rpl}  routing protocol. In addition to the simulated Z1 devices that comprise the simulated IoT network, our experimental setup includes real  physical devices; a desktop computer supporting the \textit{VSD} is connected to a Wireless Access Point (WAP) through the Ethernet ($eth0$) interface, which, in turn, is connected to a physical Android smart phone, and a laptop computer. The WAP bridges the $eth0$ interface to the WiFi medium.  A bridge is configured to connect the (simulated) 6LoWPAN network and the gateway device using the \texttt{Tunslip} utility \cite{tunslip} of Cooja.  All the devices in the network use UDP as the transport layer protocol to exchange UPnP discovery and advertisement messages.

A simulated device is configured as a resource constrained SD which provides three UPnP services, each providing two actions that can be invoked. The rest of the simulated devices are configured as CPs, which issue discovery messages to search for UPnP services provided by the SD. The smart phone (connected via WiFi) runs an application that is configured as both a CP and a SD. The laptop is configured as a traditional CP and connects via the WiFi interface. A desktop computer serves as the gateway device in the experimental network.

Our experiments evaluated the performances of basic UPnP and \model explored the energy consumption of the resource constrained IoT devices. We also analyze the throughput of the IoT network in both tradition UPnP network (WiFi) and constrained IoT UPnP network (6LoWPAN). %We did not conduct the energy consumption experiments for the traditional UPnP network because the participants of the traditional UPnP network were not as resource constrained as the IoT devices.

\subsection{Energy Consumption Analysis}
In our experimental setup, the smart phone acting as a CP that sends discovery messages to the experimental UPnP network, searching for the services provided by the IoT SD located in the IoT network. In addition, we included IoT CPs, varying their number to explore how energy consumption is impacted as the number and source of discovery messages increases. We also varied the frequency of the discovery messages to 1000 ms, 2000 ms, and 3000 ms. We chose a fixed 2 min interval between advertisement messages for a service. Each round of our experiment was 20 minutes long. We repeated each experiment 10 times and reported the averaged outcomes. To explore power consumption, our experiments leveraged the Powertrace library \cite{powertrace-api} which used the Contiki’s energy APIs to measure the power consumption of CPU and radio transceiver in the SD. 

\begin{figure}[t]
 \includegraphics[width=\columnwidth]{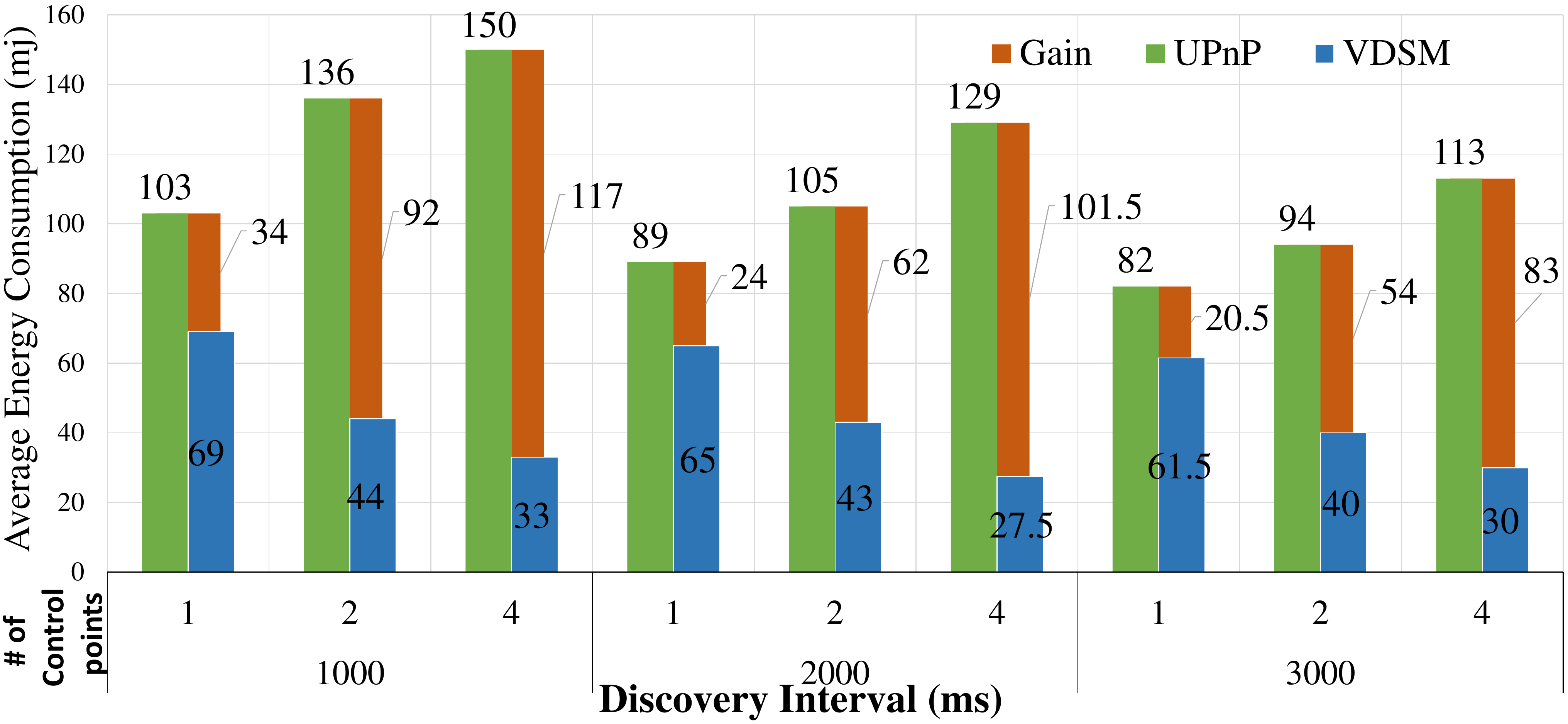}
  \begin{center}
  \caption{Gain in energy consumption UPnP vs\model}
  \label{fig:energy-exp}
  \end{center}
\end{figure}

Figure~\ref{fig:energy-exp} shows the result of the energy-consumption experiments. As expected, as we increase the number of CPs (and therefore, the number of discovery messages), we see an increase in the energy gain of the \model over basic UPnP. This is due to the service discovery delegation approach embodied in \model. With a VSD, the discovery message send by a CP is not received by the SP; instead the VSD receives and replies to the discovery messages on behalf of the SP. As a result, in \model, the SD replies to fewer messages compared to the basic UPnP. We see this benefit increase as the number of CPs who are requesting to discover services increases. For example, with discovery message frequency of 1000 ms, with 1, 2 , and 4 CPs the SD saved 34 mj, 92 mj, and 117 mj of energy, respectively. We also observed, with decreasing discovery frequency, the energy savings provided by \model decreases. As with less frequent discovery requests, the SD had less discovery reply to delegate.  
% \textcolor{red}{We also observed, with increasing discovery frequency, the energy savings provided by \model decreases. As with less frequent discovery requests, the SD had less discovery reply to delegate.}
% \footnote{JP: I don't understand the result or the rationale given here. Is this a typo and you meant "decreasing discovery frequency"?} 
With 2 participating CPs, the SD saved 67.6\%, 41\%, and 57\% of energy for 1000 ms, 2000 ms, and 3000 ms frequency, respectively. Our energy experiments show that the introduction of the VSD can dramatically reduce energy consumption due to discovery request and advertisement in resource-constrained IoT SDs in the \model compared to the basic UPnP protocol.  

\subsection{Throughput Analysis}
To measure the throughput of an UPnP, we choose a targeted CP and some participating CP. We assumed that the targeted CP already received the advertisement of the SD and know how to invoke service actions from the device and service . Th targeted CP send  action invocation requests to the SD using the control messages of UPnP with an interval of 10 seconds. Meanwhile, the participating CPs were sending discovery messages in random intervals (between 1000 ms and 3000 ms inclusive) to the SD searching for the services. As we wanted to measure the effects of the discovery reply messages and advertisements in the service action invocation requests in an UPnP network, we measure the throughout for the targeted CP using the below formula:
\[ Throughout = \frac{\text{\textit{bytes received by SD for Service Access}}}{\textit{Total End-to-end delay to send the bytes}} \]
Here, we define service access as an action invocation request (control message) or a event subscription request (eventing) sent by the targeted CP.
\begin{figure}[t]
	\centering
	
\begin{subfigure}[b]{\columnwidth}

    \includegraphics[width=\columnwidth]{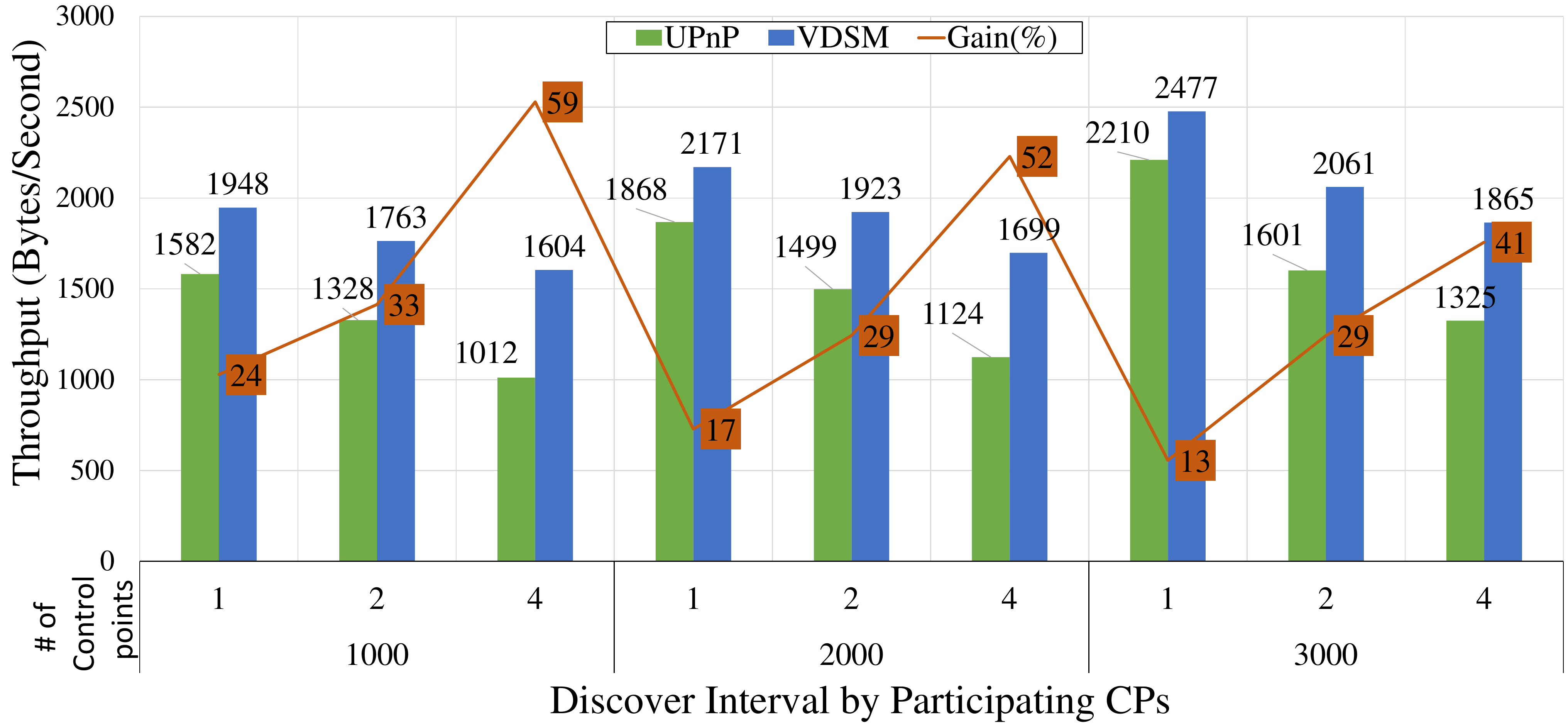}

  \caption{Throughput comparison in IoT network.}
  \label{fig:iot-throughput-exp}
  \end{subfigure}	
 \begin{subfigure}[b]{\columnwidth}

    \includegraphics[width=\columnwidth]{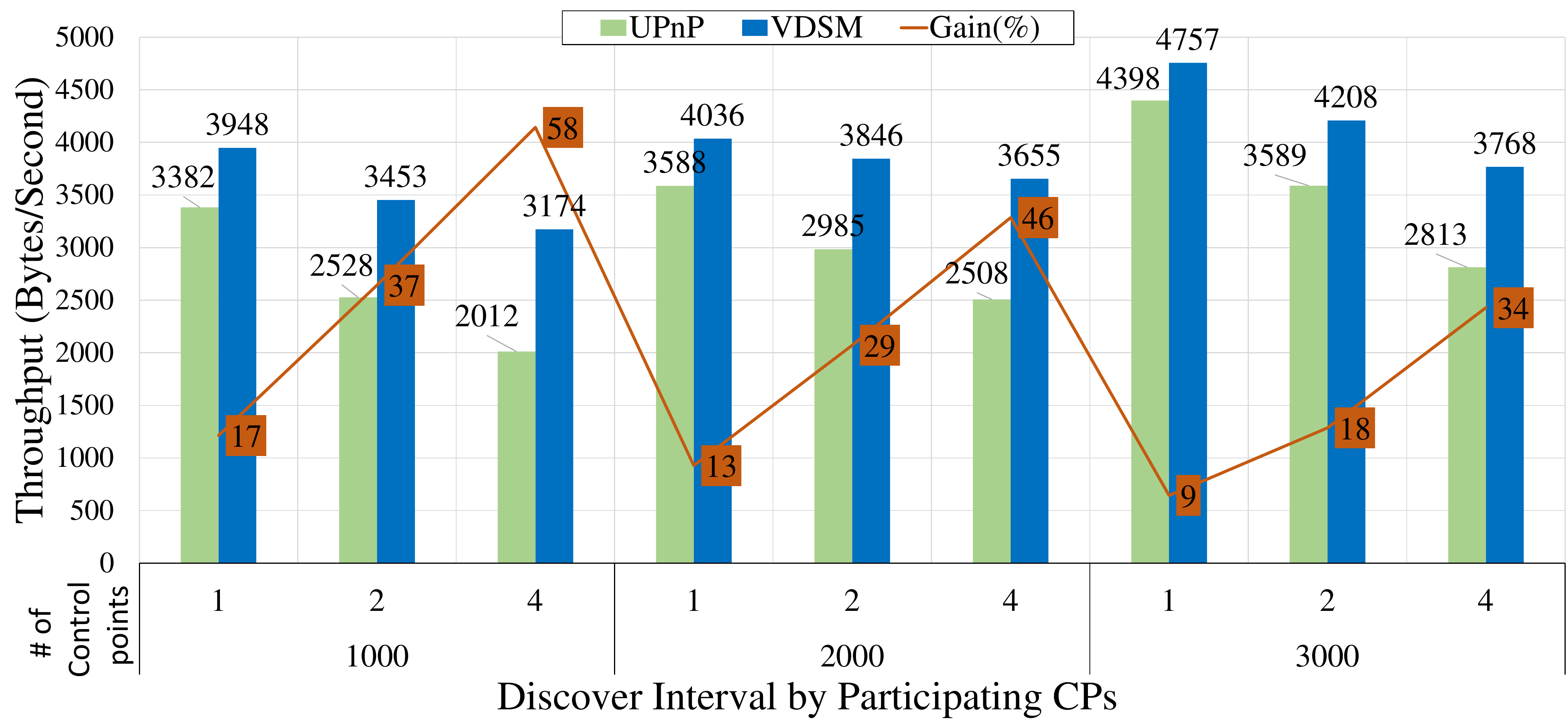}

  \caption{Through comparison in WiFi network.}
  \label{fig:wifi-throughput-exp}
  \end{subfigure}

	\caption{Throughput comparison of \model vs  UPnP network.}
	\label{fig:throughput-exp}
\end{figure}
In the resource-constrained IoT network experiments, the IoT devices act as SDs. The smart phone acts as a CP. Like in the previous experiments, we varied the number of IoT CPs, which sent discovery messages, and varied the intervals at which discovery messages were sent by the CPs to the IoT SD. Figure~\ref{fig:iot-throughput-exp} shows the result of these experiments. In the traditional UPnP network experiments, we used the smart phone as the SD and the laptop as the targeted CP. Different number of IoT CPs also took part in the experiment by sending discovery message to the smart phones. Figure~\ref{fig:wifi-throughput-exp} shows the result of the experiments for traditional UPnP network.

For both traditional and resource-constrained IoT UPnP networks, \model outperforms the basic UPnP protocol. In this \model, the VSD served the discovery replies that enabled the SDs to serve the service action invocation requests quickly. We also observed that as the frequency of the discovery message increases, the throughput decreases in both UPnP and\model. However, throughput decreases in \model at a slower rate  than for the basic UPnP. We also find that the throughput gain for \model is positively correlated with the number of participating CPs; as more CPs send more discovery messages, the SD offloads more work to the VSD, which results in increased throughput. 

\subsection{Comparative Discussion}
Yiqin et al. \cite{yiqin2009home} proposes a UPnP based networking solution to monitor and
control the home appliances and smart devices by the remotes, smart phones, laptops etc in a home network. Arunachalam et al. \cite{arunachalam2017extending} proposes an extension of UPnP to improve the interoperability among heterogeneous devices in IoT heavy home networks. The authors introduce an UPnP Application Architecture along with the UPnP application template and UPnP  service template to develop applications that run on heterogenous devices. The proposed extension basically enhance the device interoperability of UPnP to application interoperability. In \cite{mitsugi2011bridging}, researchers propose new  Constrained Application Protocol (CoAP)~\cite{coap} methods so that the UPnP services offered by a constrained network can be discovered via a CoAP/UPnP bridge co-located on the constrained network gateway. In the proposed method, UPnP messages are translated to the extended CoAP methods at the bridge and vice versa. Another similar approach~\cite{kim2011seamless} is proposed to bridge the Zigbee and the UPnP  leveraging the low energy footprint of the Zigbee protocol to reduce the energy consumption of the UPnP devices in a constrained network. Researchers \cite{8669421} propose a new service discovery protocol to make the UPnP service discovery more efficient and compact in IoT based IPv6 home networks. This extension take advantage of the improved design of IPv6 to build a new  service discovery method that increase the efficiency of transmission.  

The prior works do not recognise that the resource constrained IoT devices can reside with resource rich devices. And the overhead of performing UPnP service advertisement and discovery can be offloaded to resource-heavy counter parts of the constrained networks. Our proposed scheme achieves better energy efficiency by delegating the advertisement and handling the discovery requests in resource rich members of the constrained UPnP enabled IoT network.

%% file: conclusion-future-work.tex
s\section{Concluding Remarks}
This paper proposes an extension of the basic UPnP protocol that delegates the service advertisements and discovery requests replies from the resource-constrained IoT devices to the resource-rich elements of an UPnP-enabled IoT network. The proposed scheme leverages the insight that the resource-limited IoT devices are often co-located with resource-rich neighbours that have the potential to carry out the UPnP tasks on behalf of the IoT devices. The prototype-based evaluation shows that the proposed extension outperforms the conventional UPnP in terms of energy consumption and network throughput.

In the given solution, we have proposed the VSDM to delegate advertisement and discovery related tasks from IoT service devices. We have argued that the use of multiple VSDs in the delegation process achieve more flexibility in the implementation. In future, it would be  worth investigating, the delegation among a number of VSDs that can be optimized by distributing the delegated tasks to get improved performances. The trade-off between the number of VSDs placement and performance in an UPnP-based IoT network can also be subjected to future works.
% Besides, the proposed model gives a one-to-one relationship between SD and VSD. But adopting one-to-many relationship from SD to VSD to get a fault resilience behavior in task delegation in UPnP IoT network also a exciting directions for future researchers. Moreover, addressing the security aspects of UPnP service discovery and advertisement is highly desirable. The delegating expensive authentication and authorization tasks required in security features can be worthy research.     

%% file: acknowldgement.tex
\section*{Acknowledgements}
%This material is based upon work  supported by the US National Science Foundation (NSF) under Grant No. CNS-1828363.
This research was  supported in part by the US National Science Foundation (NSF) under Grant No. CNS-1828363 and in part the Sejong University research faculty program under the Grant No. 20192021.